\documentclass[aps,prd,showpacs,twocolumn,preprintnumbers]{revtex4}

\usepackage{amsfonts}
\usepackage{amsmath}
\usepackage{amssymb}
\usepackage{bm}
\usepackage{dcolumn}
\usepackage{epsfig}
\usepackage{graphicx}
\usepackage{graphics}
\usepackage[latin1]{inputenc}
\usepackage{latexsym}
\usepackage{rotating}
\usepackage{hyperref}

\newcommand\be{\begin{equation}}
\newcommand\ba{\begin{eqnarray}}
\newcommand\ee{\end{equation}}
\newcommand\ea{\end{eqnarray}}

\begin{document}

\title{
The Barbero-Immirzi Parameter as a Scalar Field: \\ K-Inflation from Loop Quantum Gravity?
}

\author{Victor Taveras} 
\email{victor@gravity.psu.edu}

\author{Nicol\'as Yunes}
\email{yunes@gravity.psu.edu}

\affiliation{Institute for Gravitation and the Cosmos,
  Department of Physics, The Pennsylvania State University, University
  Park, PA 16802, USA}

\begin{abstract}
  
We consider a loop-quantum gravity inspired modification of general relativity, where the Holst action is generalized by making the Barbero-Immirzi (BI) parameter a scalar field, whose value could be dynamically determined. The modified theory leads to a non-zero torsion tensor that corrects the field equations through quadratic first-derivatives of the BI field. Such a correction is equivalent to general relativity in the presence of a scalar field with non-trivial kinetic energy. This stress-energy of this field is automatically covariantly conserved by its own dynamical equations of motion, thus satisfying the strong equivalence principle. Every general relativistic solution remains a solution to the modified theory for any constant value of the BI field. For arbitrary time-varying BI fields, a study of cosmological solutions reduces the scalar field stress-energy to that of a pressureless perfect fluid in a comoving reference frame, forcing the scale factor dynamics to be equivalent to those of a stiff equation of state.  Upon ultraviolet completion, this model could provide a natural mechanism for k-inflation, where the role of the inflaton is played by the BI field and inflation is driven by its non-trivial kinetic energy instead of a potential.
   
\end{abstract}

\preprint{IGC-08/7-1}



\maketitle

\section{Introduction}

The failure of general relativity (GR) to explain the nature of spacetime and cosmological singularities begs for a completion of the theory. One path toward this completion is the unification of GR and quantum mechanics, through the postulate that spacetime itself is discrete -- loop quantum gravity (LQG)~\cite{Ashtekar:2004eh,Thiemann:2001yy,Rovelli:2004tv}. This formalism is most naturally developed within the first-order approach~\cite{Romano:1991up} in terms of a generic connection and its conjugate electric field.  When cast in these new variables, the quantization of the Einstein-Hilbert action resembles that of quantum electrodynamics, and thus, tools from field and gauge theories can be employed. 

Currently, two versions exist of the connection variables: a selfdual $SL(2,\mathbb{C})$ Yang-Mills-like connection and a real $SU(2)$ connection. The first kind is the so-called Ashtekar connection, which was the first employed to develop LQG and which must satisfy some reality conditions~\cite{Ashtekar:1991hf}. The second type is the so-called Barbero connection and it was constructed to avoid these reality conditions~\cite{Barbero:1994ap}. Both the Ashtekar or Barbero formalisms can be obtained directly from the so-called Holst action, which consists of the  Einstein-Hilbert piece plus a new term that depends on the dual to the curvature tensor~\cite{Holst:1995pc}. The Barbero-Immirzi (BI) parameter $\gamma$ arises in the Holst action as a multiplicative constant that controls the strength of the dual curvature correction. In the quantum theory, it determines the minimum eigenvalue of the discrete area and discrete volume operators~\cite{Immirzi:1996di}. 

The Holst action reduces to the Einstein-Hilbert action upon imposition of the field equations obtained through the action-principle. Variation with respect to the connection reduces to a torsion-free condition, and when this is used at the level of the action, the dual curvature piece vanishes due to the Bianchi identities. Therefore, the Holst action leads to the same dynamical field equations as the Einstein-Hilbert action, with modifications only in the quantum regime. In the presence of matter, such as fermions, the dual curvature piece does not vanish identically since the variation of the action with respect to the connection leads to a non-vanishing torsion tensor~\cite{Perez:2005pm,Freidel:2005sn}.

In this paper, we consider a generalization of GR, modified Holst gravity, where we {\emph{scalarize}} the BI parameter in the Holst action, i.e. we promote the BI parameter to a field under the integral of the dual curvature term. Allowing the BI field to be dynamical implies that derivatives of this field can no longer be set to zero when one varies the action and integrates it by parts. These derivatives generically lead to a torsion-full condition that produces non-trivial modifications to the field equations. 

Scalarization is motivated in two different ways. One such way is the study of the possible variation of what we believe to be ``universal physical constants.''  The study of models that allow non-constant couplings have a long history, one of the most famous of which, perhaps, is the so-called Jordan-Brans-Dicke theory~\cite{Jordan:1959eg,Brans:1961sx}. In this model, the universal gravitation constant $G$ is effectively replaced by a time varying coupling field, such that $\dot{G} \neq 0$ (see eg.~\cite{lrr-2006-3} and references therein for a review). Along these same lines, one could consider the possibility of a non-constant Holst coupling, where the variation could arise for example due to renormalization of the quantum theory.  Such a possibility could and does lead to interesting corrections to the dynamics of the field equations that deserve further consideration. 

Another motivation for scalarization is rooted in a recent proposal of a parity-violating correction to GR in $4$ dimensions:  Chern-Simons modified gravity~\cite{Jackiw:2003pm}. In this model, a Pontryagin density is added to the Einstein-Hilbert action, multiplied by a $\theta$-field that controls the strength of the correction. The Pontryagin density, however, can also be written as the divergence of some current (the so-called Pontryagin current). Thus, upon integrating the action by parts, the Chern-Simons correction can be thought of as the projection of the Pontryagin current along the {\emph{embedding}} vector $\partial\theta$. 
Modified Holst gravity can also be interpreted as the embedding of a certain current along the direction encoded in the exterior derivative of the BI field. In this topological view, $\partial \gamma$ acts as an embedding coordinate that projects a certain current, given by the functional integral of the dual curvature tensor.

We shall here view modified Holst gravity as a viable, falsifiable model that allows us to study the dynamical consequences of promoting the BI parameter to a dynamical quantity. This model is a formal enlargement or deformation of GR in the phase space of all gravitational theories, in which the BI scalar acts as a dynamical deformation parameter, where $\gamma = \textrm{const.}$ corresponds to GR. Arbitrarily close to this fixed point, one encounters deviations from GR, which originate from a modified torsional constraint that arises upon variation of the modified Holst action with respect to the spin connection. We shall explicitly solve this constraint to find that the torsion and contorsion tensors become proportional to first derivatives of the BI field. In the absence of matter, we find no parity violation induced by such a torsion tensor, as opposed to other modifications of GR that do include matter~\cite{Perez:2005pm,Freidel:2005sn,Alexander:2008wi}.  

The variation of the action with respect to the tetrad yields the field equations, which differ from those of GR due to the non-vanishing contorsion tensor. The modification to the field equations are found to be quadratic in the first-derivatives of the BI field, and in fact equivalent to a scalar field stress-energy tensor with no potential and non-trivial kinetic energy. This stress energy is shown to be covariantly conserved, provided the BI field satisfies the equation of motion derived from the variation of the action with respect to this field. Since the BI field now possesses equations of motion, it is dynamically determined and not fixed {\emph{a priori}}. Moreover,  the motion of point particles is still determined by the divergence of their stress-energy tensor and unaffected by the Holst modification, allowing the modified theory to satisfy the strong equivalence principle. 

Solutions of the modified theory are also studied, both for slowly-varying and arbitrarily-fast, time-varying BI fields. Since the modification to the field equations depends on derivatives of the BI scalar, every GR solution remains a solution of the modified model for constant $\gamma$. For slowly-varying BI fields, we find that GR solutions remain solutions to the modified theory up to second order in the variation of the BI parameter, due to the structure of the stress-energy tensor. In fact, gravitational waves in a Minkowski or Friedmann-Roberston-Walker (FRW) background remain unaffected by the Holst modification, and the BI field is seen to satisfy a wave equation. For  arbitrarily-fast, time-varying BI fields, cosmological solutions are considered and the scalar field stress energy tensor induced by the Holst modification is found to reduce to that of a pressureless perfect fluid in a comoving reference frame. For a flat FRW background and in the absence of other fields, the scale factor is shown to evolve in the same way as in the presence of a stiff perfect fluid. 

Finally, an effective action is constructed by reinserting the solution to the torsional constraint into the modified Holst action, which is found to lead to the same dynamics as the full action. This effective action corresponds again to that of a scalar field with no potential but non-trivial kinetic energy. Such non-trivial kinetic terms in the action prompt the comparison of modified Holst gravity to k-inflationary models, in which the inflaton is driven not by a potential but by non-standard kinetic terms. Modified Holst gravity only contains non-trivial quadratic first-derivatives of the scalar field, which in itself is insufficient to lead to inflation in the k-inflationary scenario~\cite{ArmendarizPicon:1999rj}. However, inflationary solutions are found to be allowed provided quadratic curvature corrections are added to the modified Holst action, which are prone to arise upon a UV completion of the theory. 

The remainder of this paper deals with details that establish the results summarized above. We shall here adopt the following conventions. 
Capitalized Latin letters $I,J,\ldots = 0,1,2,3$ stand for internal Lorentz indices, while lower Greek letters 
$\mu, \nu, \ldots = 0,1,2,3$ stand for spacetime indices.
Spacetime indices are usually suppressed in favor of wedge products and internal indices. We also choose the Lorentzian metric signature
$(-,+,+,+)$ and the Levi-Civit\`{a} symbol convention $\tilde{\eta}_{0123} = + 1$, which implies $\tilde{\eta}^{0123} = -1$.
Square brackets around indices stand for antisymmetrization, such as $A_{[ab]} = (A_{ab} - A_{ba})/2$.
Other conventions and notational issues are established in the next section and in the Appendix.
%

\section{Modified Holst Gravity}

In this section we introduce modified Holst gravity and establish some notation. Let us consider the following action in first-order form:
\ba
\label{Holst}
S &=& \frac{1}{4 \kappa} \int \epsilon_{IJKL} e^{I} \wedge e^{J} \wedge F^{KL}
\nonumber \\
&+& \frac{1}{2 \kappa} \int  \bar{\gamma} \; e^{I} \wedge e^{J} \wedge  F_{IJ}
+ S_{\textrm{mat}},
\ea
where $\kappa = 8 \pi G$, $S_{\textrm{mat}}$ is the action for possible additional matter degrees of freedom, $\epsilon_{IJKL}$ is the Levi-Civit\`{a}a tensor, $e$ is the determinant of the tetrad $e^{I}$ and $e_{I}$ is its inverse. In Eq.~\eqref{Holst}, the quantity $F^{IJ}$ is the curvature tensor of the Lorentz spin connection $w^{IJ}$, while $\bar{\gamma} = 1/\gamma$ is a coupling field, with $\gamma$ the BI field. Note that the first term in Eq.~\eqref{Holst} is the standard Einstein-Hilbert piece, while the second term reduces to the standard  Holst piece in the limit $\bar{\gamma} = \textrm{const.}$ (or $\gamma = \textrm{const.}$). Also note that in the modified theory there are three independent degrees of freedom, namely the tetrad, the spin connection and the coupling field.  

Varying the action with respect to the degrees of freedom one obtains the field equations of the modified theory. Assuming that the additional matter action
does not depend on the connection and varying the full action with respect to this quantity, one obtains
\be
\label{var-w}
\epsilon_{IJKL} T^{I} \wedge e^{J} = - e_{K} \wedge e_{L} \wedge D_{(\omega)}\bar{\gamma} - 2 \; \bar{\gamma} \; T_{[K} \wedge e_{L]}, 
\ee
where  $D_{(\omega)}$ stands for covariant differentiation with respect to the spin connection and the torsion tensor is defined as $T^{I} = D_{(\omega)}e^{I}$. One can arrive at Eq.~\eqref{var-w} by noting that $\delta_{\omega}R^{KL} = D_{(\omega)}\delta\omega^{KL}$, where $\delta_{\omega}$ is shorthand for the variation with respect to the spin connection, and integrating by parts. We shall here ignore boundary contributions that arise when integrating by parts, since they shall not contribute to the scenarios we shall investigate in later sections (gravitational waves and cosmological solutions). For the case of black hole solutions, such boundary terms could modify black hole thermodynamics in the quantum theory, but this goes beyond the scope of this paper. 

The remaining field equations can be obtained by varying the modified Holst action with respect to the tetrad and the coupling field. Varying first with respect to the tetrad we find
\be
\label{var-e}
\epsilon_{IJKL} e^{J} \wedge F^{KL} = - 2 \bar{\gamma} e^{J} \wedge F_{IJ},
\ee
where again we have assumed the additional matter degrees of freedom do not depend on the tetrad. 
Varying now the action with respect to the coupling field we find
\be
\label{var-gamma}
\frac{\delta S_{\textrm{mat}}}{\delta \bar{\gamma}} = - \frac{1}{2 \kappa} e^{I} \wedge D_{(\omega)}T_{I},
\ee
where we have assumed the matter degrees of freedom could contain a contribution that depends on $\bar{\gamma}$ and thus the BI field. 

The field equations of modified Holst gravity are then Eqs.~\eqref{var-w}-\eqref{var-gamma}. Note that for any non-constant value of $\bar{\gamma}$, the Holst modification leads to a torsion theory of gravity. More interestingly, even if $\bar{\gamma} = 0$, modified Holst gravity also leads to torsion provided the derivatives of the BI parameter are non-vanishing. In fact, Eq.~\eqref{var-e} resembles the Einstein equations in the presence of matter, where the matter stress-energy is given by the covariant derivative of the torsion tensor. Such a resemblance is somewhat deceptive because the curvature tensor is {\emph{not}} the Riemann tensor, but a generalization thereof, which also contains torsion-dependent pieces. Thus, the full modified field equations can only be obtained once Eq.~\eqref{var-w} is solved for the torsion and contorsion tensors.  

\section{Torsion and Contorsion in Modified Holst Gravity}

In this section we solve for the torsion and contorsion tensors inherent to modified Holst gravity. Equation~\eqref{var-w} is difficult to solve in its standard form, so instead of addressing  it directly we shall follow the method introduced by~\cite{Perez:2005pm}. 

Let us then simplify Eq.~\eqref{Holst} in the following manner
\ba
S &=& \frac{1}{4 \kappa} \int \epsilon_{IJKL} e^{I} \wedge e^{J} \wedge e^{P} \wedge e^{Q} \frac{1}{2} F^{KL}{}_{PQ}
\nonumber \\
&+& \frac{1}{2 \kappa} \int  \bar{\gamma} \; e^{I} \wedge e^{J} \wedge e^{K} \wedge e^{L} \frac{1}{2} F_{IJKL}
+ S_{\textrm{mat}},
\label{eq1}
\\
&=& \frac{1}{8 \kappa} \int \epsilon_{IJKL} \left( - \tilde{\sigma} \right) \epsilon^{IJPQ} F^{KL}{}_{PQ}
\nonumber \\
&+& \frac{1}{4 \kappa} \int  \bar{\gamma} \left(- \tilde{\sigma} \right) \epsilon^{IJKL} F_{IJKL}
+ S_{\textrm{mat}},
\label{eq2}
\\
&=& \frac{1}{8 \kappa} \int (-4) \left( - \tilde{\sigma} \right) \delta^{[PQ]}_{KL} F^{KL}{}_{PQ}
\nonumber \\
&+& \frac{1}{4 \kappa} \int  \bar{\gamma} \left(- \tilde{\sigma} \right) \epsilon^{IJKL} F_{IJKL}
+ S_{\textrm{mat}},
\label{eq3}
\\ \nonumber  
&=& \frac{1}{2 \kappa} \int  \tilde{\sigma} \left[ \delta^{[PQ]}_{KL} F^{KL}{}_{PQ}
- \frac{\bar{\gamma} }{2}   \epsilon^{IJKL} F_{IJKL} \right] + S_{\textrm{mat}},
\ea
where $\tilde{\sigma} = d^{4}x \sqrt{-g} = d^{4}x \; e$. In Eq.~\eqref{eq1} we reinstated all indices 
of the curvature tensor following the conventions in the Appendix. In Eq.~\eqref{eq2}, we have used
the following identity:
\be
\label{identity1}
e^{I} \wedge e^{J} \wedge e^{K} \wedge e^{L} = - \tilde{\sigma} \epsilon^{IJKL},
\ee
which derives from the relation $e^{0} \wedge e^{1} \wedge e^{2} \wedge e^{3} = 1/4! \; \tilde{\eta}_{IJKL} \; e^{I} e^{J} e^{K} e^{L}$, 
where $\tilde{\eta}_{IJKL}$ is the Levi-Civit\`{a} symbol. Equation~\eqref{eq3} makes use of the $\delta-\epsilon$ relation, 
which in four-dimensions reduces to
\be
\epsilon^{IJKL} \epsilon_{IJPQ} = -4\delta^{[KL]}_{PQ} 
:= -4\delta^{[K}_{P} \delta^{L]}_{Q} =  -2\left(\delta^{K}_{P} \delta^{L}_{Q} - \delta^{L}_{P} \delta^{K}_{Q} \right) .
\ee

The modified Holst action can thus be recast as follows:
\be
\label{action2}
S = \frac{1}{2 \kappa} \int d^{4}x \; e\; p^{IJ}{}_{KL}   \; e_{I}^{\mu} e_{J}^{\nu}  F^{KL}{}_{\mu \nu},
 \ee
where the operator $p^{IJ}{}_{KL}$ is given by 
\be
p_{IJ}{}^{KL} = \delta^{[K}_{I} \delta^{L]}_{J}- \frac{\bar{\gamma}}{2} \epsilon_{IJ}{}^{KL}.
\ee
In terms of this operator, Eq.~\eqref{var-w} becomes 
\be
p^{IJ}{}_{KL} D_{(\omega)} \left(e_{I}^{\mu} e_{J}^{\nu} \right) = \frac{1}{2} e_{I}^{\mu} e_{J}^{\nu} \epsilon^{IJ}{}_{KL} D_{(\omega)} \bar{\gamma},
\ee
and after isolating the torsion tensor we obtain
\ba
\label{tor-cond1}
2 T_{[I} \wedge e_{J]}  &=&  \frac{\partial_{Q} \bar{\gamma}}{2 \bar{\gamma}^{2} + 2} \left[ \epsilon_{MNIJ} e^{M} \wedge e^{N} \wedge e^{Q} 
\right. 
\nonumber \\
&-& \left. 
2 \bar{\gamma} e_{I} \wedge e_{J} \wedge e^{Q} \right],
\ea
where we have employed the inverted projection tensor
\be
(p^{-1})_{KL}{}^{IJ} = \frac{1}{\bar{\gamma}^{2} + 1} \left( \delta_{K}^{[I} \delta^{J]}_{L} + \frac{\bar{\gamma}}{2} \epsilon_{KL}{}^{IJ} \right).
\ee

The torsion tensor can now be  straightforwardly computed by solving the torsion condition [Eq.~\eqref{tor-cond1}]
to find
\be
T^{I} = \frac{1}{2} \frac{1}{\bar{\gamma}^{2} + 1} \left[ \epsilon^{I}{}_{JKL}  \partial^{L}\bar{\gamma} + \bar{\gamma} \; \delta^{I}_{[J} \partial_{K]} \bar{\gamma}  \right] e^{J} \wedge e^{K}
\ee
This expression can be shown to solve Eq.~\eqref{var-w}, thus satisfying the field equation associated with the variation of the action with respect to the spin connection.

Before we can address the modified field equations for the tetrad fields, we must first calculate the contorsion tensor. This tensor plays a critical role in the construction of the spin curvature, correcting the Riemann curvature through torsion-full terms.  
Let us then split this connection into a symmetric, tetrad compatible piece $\Gamma^{IJ}$ and an antisymmetric piece $C^{IJ}$, called the contorsion:
\be
\omega^{IJ} = \Gamma^{IJ} + C^{IJ},
\ee
In the Appendix, we derive the relation between the contorsion and torsion tensor, so in this section it suffices  to mention that they satisfy 
\be
C_{IJK} = - \frac{1}{2} \left(T_{IJK} + T_{JKI} + T_{KJI} \right),
\ee
where here we have converted the suppressed spacetime index into an internal one with the tetrad. 

The contorsion tensor is then simply 
\be
\label{contorsion}
C_{IJ} = - \frac{1}{2} \frac{1}{\bar{\gamma}^{2} + 1} \left( \epsilon_{IJKQ} e^{K} \partial^{Q} \bar{\gamma}  - 2 \bar{\gamma} e_{[I} \partial_{J]} \bar{\gamma} \right).
\ee
One can verify that this tensor indeed satisfies the required condition $T_{IJK} = - 2 C_{I[JK]}$. 

\section{Field Equations in Modified Holst Gravity}
The field equations in modified Holst gravity are given by Eqs.~\eqref{var-w}-\eqref{var-gamma}, the first of which (the torsion condition) was already solved for in the previous section.  We are then left with two sets of coupled partial differential equations, one equation for the reciprocal of the BI field $\bar{\gamma}$ [Eq.~\eqref{var-gamma}] and ten equations for the tetrad fields [Eq.~\eqref{var-e}]. 

Let us begin with the equation of motion for $\bar{\gamma}$. We can compute the right-hand side of Eq.~\eqref{var-gamma}, by first calculating the covariant derivative of the torsion tensor. Upon contraction with a tetrad, this quantity is given by
\be
e_{I} \wedge D_{(\omega)}T^{I} = - 3 \tilde{\sigma} \frac{\bar{\gamma}}{\left(\bar{\gamma}^{2} + 1 \right)^{2}} \left(\partial  \bar{\gamma}\right)^{2} + 3  \tilde{\sigma} \frac{1}{\bar{\gamma}^{2} + 1} \square\bar{\gamma},
\ee
where $\square = D_{L} D^{L} = g^{\mu \nu} D_{\mu} D_{\nu}$ is the covariant D'Alembertian operator, $\tilde\sigma = \sqrt{-g} = e$ is the volume element 
and $\left(\partial \bar{\gamma} \right)^{2} := (\partial_{L} \bar{\gamma}) (\partial^{L} \bar{\gamma}) = g^{\mu \nu} (\partial_{\mu} \bar{\gamma}) (\partial_{\nu} \bar{\gamma})$.

The equation of motion for $\bar{\gamma}$ then becomes
\be
\label{FE-gamma}
\frac{\delta S_{\textrm{mat}}}{\delta \bar{\gamma}} = 
\frac{3\tilde{\sigma}}{2 \kappa}  \frac{\bar{\gamma}}{\left(\bar{\gamma}^{2} + 1 \right)^{2}} \left(\partial \bar{\gamma}\right)^{2} - \frac{3\tilde{\sigma}}{2 \kappa}  \frac{1}{\bar{\gamma}^{2} + 1} \square\bar{\gamma}.
\ee
Even in the absence of other matter degrees of freedom, the field equations themselves guarantee that the BI coupling be dynamical.  In a later section we shall study solutions to the equations of motion in different backgrounds that are approximate solutions to the modified field equations in the new theory.

In order to obtain the modified field equations let us contract $e^{S} \wedge$ into Eq.~\eqref{var-e}. Doing so we obtain,
\ba
\label{mod-FE1}
\bar{F}^{S}{}_{I} - \frac{1}{2} \delta^{S}_{I} \bar{F} = - \frac{\bar{\gamma}}{2} \epsilon^{SJPQ} \bar{F}_{IJPQ} - \frac{1}{4} \bar{F} e^{S} \wedge e_{I},
\ea
where note that the last term will later vanish because it is totally antisymmetric and we shall drop it henceforth. The overhead bar in Eq.~\eqref{mod-FE1} 
is a reminder that no indices have been suppressed, and thus, here $\bar{F}_{IJ} := \delta^{KL} \bar{F}^{K}{}_{ILS}$ and 
$\bar{F} = \delta^{[PQ]}_{KL} \bar{F}^{KL}{}_{PQ} = \delta^{IJ} \bar{F}_{IJ}$. 

With this notation, the modified field equations resemble that of GR, except that here
the $\bar{F}$ tensor is not the Ricci curvature but it also contains corrections due to torsion. 
Let us then decompose the curvature tensor into the Riemann curvature plus additional terms that depend on the contorsion tensor:
\be
\bar{F}^{IJ}{}_{KL} = R^{IJ}{}_{KL} + H^{IJ}{}_{KL},
\ee
where $H^{IJ}{}_{KL}$ stands for the Holst correction tensor 
\be
H^{IJ}{}_{KL} := 2D^{(\Gamma)}_{[K} C^{IJ}{}_{L]} + 2 C^{I}{}_{M[K}  C^{MJ}{}_{L]}
\ee
with $D_{(\Gamma)}$ the covariant derivative associated with the symmetric connection. We then find that Eq.~\eqref{mod-FE1} reduces to
\be
\label{mod-FE2}
G^{S}{}_{I} = - \left(H^{S}{}_{I} - \frac{1}{2} \delta^{S}_{I} H \right) - \frac{\bar{\gamma}}{2} \epsilon^{SJPQ} H_{IJPQ},
\ee
where again $H_{IJ} := \delta^{KL} H_{KILS}$ and $H = \delta^{[PQ]}_{KL} H^{KL}{}_{PQ}$. The right-hand side of Eq.~\eqref{mod-FE2}
acts as a stress energy tensor for the reciprocal of the BI scalar.

The remainder of the calculation reduces to the explicit calculation of the Holst correction tensor for the contorsion found in the previous section. 
In a sense, Eq.~\eqref{mod-FE2} is similar to the decomposition of the correction into irreducible pieces: a trace, a symmetric piece and an antisymmetric
piece. The calculation of these pieces is simplified if we first calculate the covariant derivative of the contorsion and the contorsion squared, namely
\ba
D^{(\Gamma)}_{M} C^{KL}{}_{N} &=&  - \frac{1}{2} \frac{1}{\bar{\gamma}^{2} + 1} \left\{ \left[ - \frac{2 \bar{\gamma}}{\bar{\gamma}^{2} + 1} 
\partial_{M} \bar{\gamma} \partial^{Q} \bar{\gamma}  
\right. \right.
\nonumber \\
&+& \left. \left. 
 D^{(\Gamma)}_{M}\partial^{Q} \bar{\gamma} \right]
\epsilon^{KL}{}_{NQ} + \left[ 2 \frac{\bar{\gamma}^{2} - 1}{\bar{\gamma}^{2} + 1} \partial_{M} \bar{\gamma}  
\right. \right. 
\nonumber \\
&-& \left. \left. 
 2 \bar{\gamma} D_{M}^{(\Gamma)} \right]
\delta^{[K}_{N} \partial^{L]} \bar{\gamma}\right\}
\nonumber \\
C^{K}{}_{M[Q}  C^{ML}{}_{T]} &=& \frac{1}{4} \frac{1 - \bar{\gamma}^{2}}{\left(\bar{\gamma}^{2} + 1 \right)^{2}} \left[ \left(\partial\bar{\gamma}\right)^{2} \delta^{K}_{[Q} \delta^{L}_{T]} + 2 \left(\partial^{[K} \bar{\gamma}\right)   
\right. 
\nonumber \\
&\times& \left.
\delta^{L]}_{[Q} \; \left( \partial_{T]}\bar{\gamma} \right) 
\right. 
\nonumber \\
&+& \left. 
\frac{2 \bar{\gamma}}{1 - \bar{\gamma}^{2}} \left(\partial^{[K} \bar{\gamma}\right) \epsilon^{L]}{}_{TQS} \left(\partial^{S} \bar{\gamma} \right) \right].
\nonumber \\
\ea
 With these expressions at hand, the Holst correction tensor is given by
 \ba
 H^{I}{}_{J} &=& \; _{1}H^{I}{}_{J} + \; _{2}H^{I}{}_{J}
 \nonumber \\
 \; _{1}H^{I}{}_{J} &:=& 2D^{(\Gamma)}_{[K} C^{KI}{}_{J]} 
 \nonumber \\
&=& \frac{1}{\bar{\gamma}^{2} + 1} \left\{  \frac{\bar{\gamma}^{2} - 1}{\bar{\gamma}^{2} + 1} \left[ \left(\partial_{I} \bar{\gamma}\right) \left(\partial^{J} \bar{\gamma} \right) 
+ \frac{1}{2} \delta^{I}_{J} \left(\partial{\bar{\gamma}}\right)^{2} \right]
\right. 
\nonumber \\
&-& \left. 
  \bar{\gamma} \left[ D_{J} \partial^{I} \bar{\gamma} + \frac{1}{2} \square{\bar{\gamma}} \delta^{I}_{J} \right] \right\} 
 \nonumber \\
\; _{2}H^{I}{}_{J} &:=&  2 C^{K}{}_{M[K}  C^{MI}{}_{J]}
\nonumber \\
&=& \frac{1}{2} \frac{1 - \bar{\gamma}^{2}}{\left(1 + \bar{\gamma}^{2} \right)^{2}} \left[ \left(\partial{\bar{\gamma}}\right)^{2} \delta^{I}_{J} 
- \left(\partial^{I} \bar{\gamma}\right) \left( \partial_{J} \bar{\gamma}\right) \right],
 \ea
and its trace is then simply
 \be
 H = \frac{3}{2} \frac{\bar{\gamma}^{2} - 1}{\left( \bar{\gamma}^{2} + 1 \right)^{2}} \left(\partial \bar{\gamma}\right)^{2} - \frac{ 3 \; \bar{\gamma} \; \square \bar{\gamma} }{\bar{\gamma}^{2} + 1}. 
 \ee
Finally, the antisymmetric part of this tensor is given by
\ba
\epsilon_{SJ}{}^{PQ} H^{IJ}{}_{PQ} &=& - \frac{6 \bar{\gamma}}{\left(\bar{\gamma}^{2} + 1 \right)^{2}} \left[  \left( \partial_{S} \bar{\gamma} \right) \left( \partial^{I} \bar{\gamma} \right) 
- \frac{1}{2} \left( \partial\bar{\gamma} \right)^{2} \delta^{I}_{S}\right] 
\nonumber \\
&-& \frac{2}{\bar{\gamma}^{2} + 1} \left( \delta^{I}_{S} \square \bar{\gamma} - D^{I}\partial_{S} \bar{\gamma} \right)  
 \ea

We have now all the machinery in place to compute the modification to the field equations [ie.~the right-hand side of Eq.~\eqref{mod-FE2}]. Combining
all irreducible pieces of the Holst correction tensor, we find
\be
G^{S}{}_{I} = \frac{3}{2} \frac{1}{\bar{\gamma}^{2} + 1} \left[ \left( \partial^{S} \bar{\gamma} \right) \left( \partial_{I} \bar{\gamma} \right) - \frac{1}{2} \delta^{S}_{I} \left(\partial \bar{\gamma} \right)^{2} \right], 
\ee
or in terms of spatial indices
\be
G_{\mu \nu} = \frac{3}{2} \frac{1}{\bar{\gamma}^{2} + 1} \left[ \left( \partial_{\mu} \bar{\gamma} \right) \left( \partial_{\nu} \bar{\gamma} \right) - \frac{1}{2} g_{\mu \nu} \left(\partial \bar{\gamma} \right)^{2} \right]. 
\ee
Remarkably, the second derivatives of $\bar{\gamma}$ have identically cancelled upon substitution of the solution to the torsional constraint. Perhaps even more remarkably, 
we find that modified Holst gravity is exactly equivalent to GR in the presence of an BI field with stress-energy tensor 
\be
\label{Tab}
T_{\mu \nu} = \frac{3}{2 \kappa} \frac{1}{\bar{\gamma}^{2} + 1} \left[ \left( \partial_{\mu} \bar{\gamma} \right) \left( \partial_{\nu} \bar{\gamma} \right) - \frac{1}{2} g_{\mu \nu} \left(\partial \bar{\gamma} \right)^{2} \right].
\ee
Such a stress-energy tensor is similar to that of a scalar field, except for a scalar-field dependent prefactor and the fact that $\bar{\gamma}$ obeys a more complicated and 
non-linear evolution equation [Eq.~\eqref{FE-gamma}] than the scalar field one.

\section{Solutions in Modified Holst Gravity}

Now that the field equations have been obtained, one can study whether well-known solutions in GR are still solutions in modified Holst gravity.
Formally, in the limit $\bar{\gamma} = \textrm{const.}$, all GR solutions remain solutions of the modified theory.  
If one adopts the view that derivatives of the BI field are small, then to first order in these derivatives, all standard solutions in GR also remain solutions of modified Holst gravity. 
This is because the stress energy tensor found in the previous section depends quadratically on derivatives of the BI field, and thus, can be neglected to first order. 

Perturbations of standard solutions that solve the linearized Einstein equations, however, need not in general be also solutions to the linearized modified Holst field equations. 
For example, perturbations of the Schwarzschild spacetime will now acquire a source that depends on the BI field. This source could in turn modify the gravitational wave 
emission of such perturbed spacetime, and thus, the amount of energy-momentum carried by such waves.

Exact solutions to the modified field equations are difficult to find, due to the non-trivial coupling of the BI scalar to all metric components. We can however study some
of the perturbative features of this theory to first order. In the next subsections we shall do so for spacetimes with propagating  gravitational waves and FRW metrics. 

\subsection{Gravitational Waves and Other Approximate Solutions}

Let us begin by assuming a flat background metric with a gravitational wave perturbation
\be
g_{\mu \nu} = \eta_{\mu \nu} + h_{\mu \nu}.
\ee
In the limit $\bar{\gamma} = \textrm{const.}$, all solution of the Einstein equations are also solutions of the modified Holst equations, and thus, Minkowski is also a solution. The Minkowski metric then is the background solution we shall employ.  

Let us now concentrate on the first-order evolution of $\bar{\gamma}$ in a Minkowski background. Since we are assuming the BI field varies slowly, terms quadratic in $\partial \bar{\gamma}$ can be neglected, and the equation of motion for $\bar{\gamma}$ becomes 
\be
\left(- \partial_{t}^{2} + \partial_{k} \partial^{k} \right) \bar{\gamma} = 0,
\ee
whose solution is
\be
\bar{\gamma} = \bar{\gamma}_{C} \cos\left(\omega t - k_{i} x^{i}\right) + \bar{\gamma}_{S} \sin\left(\omega t - k_{i} x^{i}\right),
\ee
where $\bar{\gamma}_{C,S}$ are constants of integration, while $\omega^{2} = k_{i} k^{i}$ is the dispersion relation, with $\omega$ the angular velocity and $k_{i}$ the wave-number vector in the direction of propagation. 

Now that the evolution of the BI field has been determined to first order, one can study first-order gravitational wave perturbations about Minkowski spacetime. 
In doing so, the modified field equations become
\be
G_{\mu \nu}[h_{\sigma \delta}] = {\cal{O}}(\omega/\Omega)^{2},
\ee
where $\Omega$ is the gravitational wave frequency. Note that the left-hand side stands for differential operators acting on the metric perturbation ({\emph{i.e.} in the Lorentz gauge, this operator would be the flat-space Laplacian), while the right-hand side stands for terms of second-order in the variation of the BI field. Thus, in modified Holst gravity, gravitational waves obey the same wave equation as in GR, to leading order in the variation of the BI field. 

The equation of motion for $\bar{\gamma}$ can also be solved exactly in a flat background, namely
\be
\label{wave}
\bar{\gamma} = \bar{\gamma}_{0} \ln \left( 1 + k_{\mu} x^{\mu} \right)
\ee
where $\bar{\gamma}_{0}$  and $k^{\mu}$ are constants of integration. One can check that for $c_{\mu} x^{\mu} \ll 1$ one recovers the linearized version of the wave solution presented above.
Of course, if all orders in the derivatives of the BI field are retained, gravitational wave perturbations will be modified, but then one must treat the coupled system simultaneously. Such a study is beyond the scope of this paper.

The result presented above is of course not dependent on the background chosen. For example, let us consider a Friedmann-Robertson-Walker (FRW) background in comoving coordinates
\be
\label{FRW}
ds^{2} = a(\eta) \left(-d\eta^{2} + d\chi_{i} d\chi^{i} \right),
\ee
where $a(\eta)$ is the conformal factor, $\eta$ is conformal time and $\chi^{i}$ are comoving coordinates. As before, to zeroth order in the derivatives of the BI field, the FRW metric remains an exact solution of the modified theory. Neglecting quadratic first derivatives of the BI field, its evolution is still governed by a wave equation, but this time about an FRW background:
\be
-\partial_{\eta}^{2}{\bar{\gamma}} - 2 {\cal{H}} \partial_{\eta} \bar{\gamma} + \partial_{i} \partial^{i} \bar{\gamma} = 0,
\ee
where ${\cal{H}} := \partial_{\eta} a/a$ is the conformal Hubble parameter. The solution to this equation is still obviously a wave, with comoving angular velocity and wavelength, {\emph{i.e.}} Eq.~\eqref{wave} with $k \to \tilde{k}= a(\eta) k$ and $\omega \to \tilde{\omega} = a(\eta) \omega$. 

The argument presented above can be generalized to other exact solutions. For example, the Schwarzschild and the Kerr metrics remain exact solutions to the modified Holst field equations to zeroth order in the derivatives of the BI field. In turn, $\bar{\gamma}$ is constrained to obey a wave equation in these background, neglecting its quadratic derivatives. This field would then source corrections to the background that would appear as modifications to the perturbation equations, but we shall not study these perturbations here.

\subsection{Cosmological Solutions}
Let us now consider the evolution of the universe in modified Holst gravity. Let us then consider the FRW line-element 
in cosmological non-comoving coordinates
\be
ds^{2} = - dt^{2} + a^{2}(t) \left[ \frac{dr^{2}}{1 - k r^{2}} + r^{2} \left(d \theta^{2} + \sin^{2}(\theta) d\phi^{2} \right) \right],
\ee
where $a(t)$ is the scale factor, $t$ is cosmological time and $k$ is the curvature parameter. 

First, let us study the evolution of $\bar{\gamma}$ in this background. In GR, the evolution of any cosmological stress-energy
is given by the divergence of $T_{\mu \nu}$, namely $\nabla_{\mu} T^{\mu}{}_{\nu}$. The zeroth component of this equation
is usually used to determine the scale-factor dependance of the energy density, once an equation of state is posed. In modified
Holst gravity, we find that energy conservation is automatically guaranteed, provided $\bar{\gamma}$ satisfies its own equation 
of motion [Eq.~\eqref{FE-gamma}], which in the absence of exterior source it reduces to 
\be
\square \bar{\gamma} = \frac{\bar{\gamma}}{\bar{\gamma}^{2} + 1 } \left(\partial \bar{\gamma} \right)^{2}.
\ee

In order to make progress, we shall assume that the BI field depends only on time, such that the equation of motion of its reciprocal reduces to
\be
\ddot{\bar{\gamma}} + 3 H \dot{\bar{\gamma}} = \frac{\bar{\gamma}}{\bar{\gamma}^{2} + 1} \dot{\bar{\gamma}}^{2},
\ee
where overhead dots stand for partial derivatives with respect to cosmological time and $H : = \dot{a}/a$. This equation can be solved 
exactly to find
\be
\label{gamma-sol}
\frac{\dot{\bar{\gamma}}}{\left( 1 + \bar{\gamma}^{2} \right)^{1/2}} = \frac{L_{0}^{2}}{a^{3}},
\ee
where $L_{0}$ is a constant of integration needed for dimensional consistency. Equation~\eqref{gamma-sol}
can be inverted to render
\be
\bar{\gamma}(t) =  \sinh {\cal{A}},
\ee
where we have defined
\be
\label{Aoft}
{\cal{A}}(t) := \int \frac{L_{0}^{2}}{a^{3}(t)} dt, 
\ee
which contains a hidden constant of integration.
We see then that the BI field depends on the integrated history of the inverse volume element of spacetime. Naturally, as 
spacetime contracts (near the spacelike singularity where $a(t) \to 0$), $\bar{\gamma} \to \infty$ and the BI scalar tends to zero.

Let us now return to the modified field equations. Due to the symmetries of the background, there are only two independent 
modified field equations, namely
\ba
\label{FE1}
&& - 3 \frac{\ddot{a}}{a} = \frac{3}{2} \frac{\dot{\bar{\gamma}}^{2}}{\bar{\gamma}^{2} + 1}, 
\\
\label{FE2}
&&
\frac{\ddot{a}}{a} + 2 \left(\frac{\dot{a}}{a} \right)^{2} + 2 \frac{k}{a^{2}} = 0.
\ea
We can simplify Eq.~\eqref{FE1} with both Eqs.~\eqref{FE2} and~\eqref{gamma-sol} to find
\be
\label{FE3}
\left(\frac{\dot{a}}{a}\right)^{2} = \frac{L_{0}^{4}}{4 a^{6}} - \frac{k}{a^{2}}.
\ee

Equations~\eqref{FE2} and~\eqref{FE3} are the only two independent modified field equations and they reduce to the Raychaudhuri and Friedmann 
equations respectively in the limit $\bar{\gamma} = \textrm{const.}$ The flat ($k=0$) solution to the Holst-modified Friedmann-Raychaudhuri equations is simply
$a \propto L_{0}^{2/3} (t - t_{0})^{1/3}$, where $t_{0}$ is an integration constant, associated with the classical singularity.

Interestingly, one can now reinsert this solution into Eq.~\eqref{Aoft} to study the temporal behavior of the BI scalar. Doing so, one finds that $\bar{\gamma} \propto [(t-t_{0})/t_{1}]^{2/3} - [(t-t_{0})/t_{1}]^{-2/3}$, 
where $t_{1}$ is the hidden constant of integration of Eq.~\eqref{Aoft}, which is fixed via initial conditions on $\bar{\gamma}$.  
Such a solution implies that as $t \to t_{0}$ or $t \to \infty$, $\bar{\gamma} \to \infty$, which forces the BI scalar $\gamma$ to asymptotically approach zero.

Such results, however, are at this point premature since modified Holst gravity is a {\emph{classical}} theory 
and one must analyze its quantization more carefully to determine what the $\bar{\gamma}$ field represents in terms of the spectrum of quantum geometric operators.
If we make the naive assumption that this field plays the same role in the quantized modified theory as in LQG, then in the infinite future limit $t \to \infty$, 
the spectrum of quantum geometric operators would become continuous. Surprisingly, in the infinite-past limit $t \to t_{0}$, the spectrum of geometric operators
also approaches continuity, which could indicate that the BI scalar becomes asymptotically free. 

The Holst modification with time-dependent BI field is then equivalent to GR in the presence of a perfect fluid. The stress-energy
tensor of such fluids is given by $T_{\mu \nu} = \left( \rho + p \right) u_{\mu} u_{\nu} - p g_{\mu \nu}$, where $p$ is the pressure, $\rho$ is
the energy density and $u_{\mu}$ is the $4$-velocity of the fluid. In this case, the Holst modification is equivalent to a pressureless perfect
fluid in a comoving reference frame $u_{\mu} = [-1,0,0,0]$, with energy density
\be
T_{00} = \rho = \frac{3L_0^2}{4 \kappa a^{6}}.
\ee
Such a stress energy energy in fact also leads to the same scale-factor evolution as a pressureful perfect fluid in a comoving reference frame 
with equation of state $p = w \rho$ and  $w = + 1$ . Such an equation of state is called {\emph{stiff}} in the literature. 

\section{Effective Action and Inflation}
\label{k-inflation}

The structure of the torsion and contorsion tensors remind us of the Klein-Gordon scalar field. For this reason, it is interesting to study the correction to the effective action obtained by reinserting these tensors into Eq.~\eqref{action2}. In doing so, one obtains
\ba
\label{effective-action}
S_{\textrm{eff}} &=&  \frac{1}{2 \kappa} \int \tilde{\sigma} \left[ R - \frac{3}{2} \frac{1}{\bar{\gamma}^{2} + 1} \left(\partial \bar{\gamma} \right)^{2} \right],
\ea
where again we see that the second derivatives have identically vanished. In general, the insertion of the solution to the torsion
condition into the action and its variation to obtain field equations need not commute. In this case, however, they do as one can trivially
check by varying Eq.~\eqref{effective-action} with respect to the metric. Similarly, from this effective action one can recompute the stress-energy tensor of $\bar{\gamma}$ 
to obtain Eq.~\eqref{Tab}. 

Non-trivial kinetic terms in the action, similar to those in Eq.~\eqref{Tab}, are the pillars of the k-inflationary model. 
In this model, inflation and the inflaton field are driven by such terms, instead of a potential. More precisely,  Ref.~\cite{ArmendarizPicon:1999rj} considers the following action:
\be
S_{\textrm{k}} = \frac{1}{2 \kappa} \int d^{4}x \sqrt{-g}  \left[ R - \kappa K(\psi) \left( \partial \psi \right)^{2}- \frac{\kappa}{2} L(\psi) \left(\partial \psi \right)^{4} \right],
\ee
where $\psi$ is the inflaton, while $K(\psi)$ and $L(\psi)$ are non-trivial arbitrary functions of the scalar field $\psi$. Ref.~\cite{ArmendarizPicon:1999rj} shows that this modified 
action is equivalent to GR with a perfect fluid, whose stress energy tensor $T_{\mu \nu} = \left( \rho + p \right) u_{\mu} u_{\nu} - p g_{\mu \nu}$ and its energy density and pressure
are given by 
\ba
\rho &=& \frac{1}{2} K(\psi) \left( \partial \psi \right)^{2} + \frac{3}{4} L(\psi) \left( \partial \psi\right)^{4},
\nonumber \\
p &=& \frac{1}{2} K(\psi) \left( \partial \psi\right)^{2} + \frac{1}{2} L(\psi) \left(\partial \psi \right)^{4},
\ea
with four-velocity
\be
u_{\mu} = \frac{1}{\sqrt{\left(\partial \psi \right)^{2}}} \partial_{\mu} \psi,
\ee
and with $\dot\psi > 0$. Inflation then arises provided $w = p/\rho = -1$, which corresponds to 
\be
\frac{K}{L} = - \left( \partial \psi \right)^{2}.
\ee
One then discovers that if non-trivial quadratic and quartic kinetic terms are present in the action, inflation can arise naturally without the presence of a potential.

The k-inflationary scenario can be compared now to modified Holst gravity. Doing so, one finds that the modified Holst contribution to the effective action is 
equivalent to the one considered in \cite{ArmendarizPicon:1999rj}, where, modulo a conventional overall minus sign  
\be
K =  \frac{3}{2 \kappa} \frac{1}{\bar{\gamma}^{2} + 1},
\qquad L = 0.
 \ee
Note that the functional $K$ is always positive, provided the BI field is real. If $\bar{\gamma}$ is complex (which is allowed provided $\bar{\gamma} \neq i$), then the $K$ functional 
could in fact change signs.

One is thus tempted to arrive at the perhaps surprising identification of the BI field as the inflaton of early cosmology. However, modified Holst gravity as analyzed here (without external potential contributions) is not sufficient to lead to an inflationary solution. One has already seen this in the previous section, where we found that $a(t) \propto t^{1/3} \neq e^{H t}$. In other words, since $L=0$ the energy density of the analog perfect fluid would be equal to its pressure, thus leading to a so-called ``hard'' or ``''stiff``'' equation of state and $a(t) \propto t^{1/3}$. 

Two paths can lead to inflation in modified Holst gravity. The first path is to include a potential or kinetic contribution for the BI field to the action or matter Lagrangian density (the $S_{\textrm{mat}}$ considered earlier). The obvious choice would be to simply add a quartic term of the form $N(\bar{\gamma}) \left(\partial \bar{\gamma}\right)^{4}$. Another less trivial possibility would be to include a term of the form 
\be
S_{\textrm{mat}} = \frac{1}{2} \int \epsilon_{IJKL} e^{I} \wedge e^{J} \wedge F^{KL }\left[ \left( \partial \bar{\gamma}\right)^{2}  + V(\bar{\gamma}) \right],
\ee
such that the full Lagrangian density became ${\cal{L}} = {\cal{L}}_{EH} \left[1 + \left(\partial\bar{\gamma}\right)^{2} + V(\bar{\gamma}) \right]$. Since the Einstein-Hilbert Lagrangian density contains non-trivial kinetic terms, such an additional kinetic piece would lead to quartic first derivatives and thus non-vanishing $L(\psi)$. 

Another much more natural route to produce quartic terms that does not involve adding arbitrary potential or kinetic contributions to the action is the inclusion of higher-order curvature corrections to the action. The modified Holst action corrects GR at an infrared level, without producing ultraviolet corrections. However, the effective quantum gravitational model represented by modified Holst gravity might require UV completion, just as string theory does.  
In string theory, such completion arises naturally in the form of effective Gauss-Bonnet and Chern-Simons terms. Such terms are topological in $4$-dimensions and are thus usually integrated by parts and the boundary contribution set to zero.  However, in modified Holst gravity, such terms will generically be non-vanishing. For example, a Ricci scalar squared correction to the action would lead to a new term in the effective action of the form
\ba
S_{\textrm{eff}}^{R^{2}} &=&  \frac{1}{2 \kappa} \int d^{4}x \sqrt{-g} \frac{9}{4} \frac{1}{\left(\bar{\gamma}^{2} + 1\right)^{2}} \left(\partial \bar{\gamma}\right)^{4} 
\ea
With this correction, the $L(\psi)$ function is not vanishing and in fact reduces to 
\be
L = - \frac{9}{2 \kappa} \frac{1}{\left(\bar{\gamma}^{2} + 1 \right)^{2}}. 
\ee
The ratio of functionals then becomes
\be
\frac{K}{L} = - \frac{1}{3} \left(1 + \bar{\gamma}^{2} \right),
\ee
which could generically lead to inflation. 
We thus conclude that although plain modified Holst gravity does not lead directly to inflation, it does allow for k-type inflation 
provided UV motivated, dimension-four corrections to the modified Holst action, such as a Gauss-Bonnet one, are also included in the action. 

\section{Conclusions}

We have studied an LQG-inspired generalization of GR, where the Holst action is modified by promoting the BI parameter to a dynamical scalar field. Three sets  of field equations were obtained from the variation of the action with respect 
to the degrees of freedom of the model. The first one is a non-linear, wave-like equation of motion for the reciprocal of the BI field, obtained from the variation of the 
action with respect to this field. The second one is a torsional constraint, obtained from the variation of the action with respect to the spin connection, 
which forces the spin connection to deviate from the Christoffel one. The third set corresponds to the modified field equations (a modification to 
the Einstein equations), obtained by varying the action with respect to the metric. 

The torsional constraint was found to generically lead to Riemann-Cartan theory, with a torsion-full
connection that we calculated explicitly in terms of derivatives of the BI field. From this torsion tensor, we computed the contorsion
tensor, which allowed us to calculate the correction to the curvature tensor. Once this correction was obtained, we found explicit expressions
for both the equation of motion for the reciprocal of the BI field as well as the modified field equations. The structure of the latter was in fact found equivalent to GR in the presence 
of a scalar field stress-energy tensor. This tensor was then seen to be covariantly conserved in the modified theory via the equation
of motion of the BI field, thus satisfying the strong equivalence principle.

Typically the value of the BI parameter $\gamma=\bar{\gamma}^{-1}$ is determined by black hole thermodynamics and takes the value $\gamma \simeq 0.24$. However, 
in modified Holst gravity the BI parameter is determined dynamically via its own equation of motion. In this sense, $\bar{\gamma}$ can take on an infinite
number of values in our universe (the space of solutions of $\bar{\gamma}$ is infinite-dimensional) and its precise value depends on the solution
to a coupled system of partial differential equations for $\bar{\gamma}$ and the metric. For instance, int he cosmological
context discussed in Sec. V B neglecting backreaction and for a BI scalar that is isotropic and homogeneous, the solution
we found for $\gamma$ approaches $0$ and not $0.24$ as in the black hole case.  Therefore, in this context, one would
need to introduce a suitable effective potential for $\bar{\gamma}$ to drive it to the black hole value.  We have
here only discussed the possibility of such a relaxation mechanism for the BI scalar in modified Holst gravity, but
much more work remains to be done to understand full the nonlinear behavior of $\gamma$ and to explain the inclusion of
an effective potential.  

Solutions were next studied in the modified theory. Since the correction to the field equations is in the form of quadratic first-order derivatives of 
$\bar{\gamma}$, all solutions of GR are also solutions to the modified theory if these derivatives are treated as small in some well-defined sense.
Gravitational wave perturbations about a Minkowski and FRW background were also studied and found to still be solutions of the modified theory
without any additional modifications. The reciprocal of the BI field in such backgrounds was seen to perturbatively satisfy the wave equation. 

Cosmological solutions were also investigated in the modified theory for an FRW background. The equations of motion for the reciprocal of the BI field were solved
exactly to find hyperbolic sinusoidal solutions. The modified Friedmann equations were then derived and solved to find a scale factor evolution corresponding
to that of a stiff equation of state. In fact, in an FRW background and for a time-like BI field, the modified theory was found equivalent to GR in the presence of a
pressureless perfect fluid in a comoving reference frame.  

Finally, an effective action was derived by inserting the solution to the torsional constrained into the modified Holst action. The effective action was found to be
equivalent to the standard kinetic part of a scalar field action, with a non-trivial prefactor. Such an action was then compared to the ones studied in 
the k-inflationary model, where the inflaton is driven by such non-trivial kinetic terms. As considered here, modified Holst gravity is insufficient to drive inflation, 
since the BI field is found to be too stiff. However, upon UV completion, quartic kinetic terms should naturally arise due to torsion contributions 
that are quadratic in the modified theory. The combination of such non-trivial quadratic and quartic kinetic terms could generically allow for inflationary fixed 
points in the phase space of solutions. 

Whether such inflationary solutions are truly realized remains to be studied further, but such a task is difficult
on many fronts. First lack of a UV-complted modified Holst gravity theory forces one to draw physical
inspiration from UV completions in string theory, such as Gauss-Bonnet or Chern-Simons like terms.  The inclusion
of a new Gauss-Bonnet term would require the addition of three new terms to the modified Holst action, including
quadratic curvature tensor pieces, which would render the new equations of motion greatly nonlinear. The solution
to this new system would thus necessarily have to be fully numerical and also raises the questions about the 
proper choice of intitial conditions.

Even if such a UV completion leads to a tractable system and a solution were found, its mere existence is
not sufficient to render the model viable as an inflationary scenario. One would necessarily also have to study
the duration of the inflationary period (the number of e-folds), the spectrum of perturbations, and other
tests that the standard inflationary model passes.  This paper lays the foundations for a new set of ideas
that could potentially tie together phenomoligical k-inflationary scenarios to quantum gravitational foundations.
The tools developed here will hopefully allow researchers to consider this model more carefully and finally contrast
it with experimental data.

\acknowledgments We would like to thank Abhay Ashtekar for suggesting this project and guiding us through it, providing great suggestions along the way. We are also indebted to Andrew Randono for helping us understand 
Einstein-Cartan theory and the non-trivialities of different conventions in the first-order formalism, as well as to  Stephon Alexander for pointing out the relation between k-inflation and modified Holst gravity. We would finally like to acknowledge Ben Owen and Abhay Ashtekar for their continuous support.

We acknowledge the support of the Center for Gravitational Wave
Physics, which is funded by the National Science Foundation under
Cooperative Agreement PHY 01-14375. Some calculations used the computer algebra
systems MAPLE (in combination with the GRTensorII package~\cite{grtensor}). NY acknowledges support
from the Eberly College of Science. VT acknowledges support from the Alfred P. Sloan Foundation, and the Eberly College of Science.

\appendix
\section{First Order Formalism: Conventions and Notation} 

In this appendix we establish the notation for the first order formalism used in this paper. Let us first note that
all spacetime indices are suppressed, and if reinstated, they are to be added after the internal
ones. It then follows that the tetrad $e^{I}$ and the spin connection $\omega^{KL}$ are $1$-forms on the base manifold,
while the curvature tensor associated with it, $F^{KL}$, is a $2$-form on the base manifold.

Spacetime indices are reinstated through wedge product operators, where the latter are defined by the operation
\be
\left(A \wedge B \right)_{\mu \nu} := \frac{\left(p + q\right)!}{p! q!} A_{ [\mu_{1} \ldots \mu_{p}} B_{\nu_{1} \ldots \nu_{q}]}
\ee
with $A$ and $B$ $p$- and $q$-forms respectively. Note that the wedge product satisfies the following chain rule
\be
D_{(\omega)} \left( A\wedge B \right) = \left(D_{(\omega)} A\right) \wedge B + (-1)^{q} A \wedge \left(D_{(\omega)} B \right),
\ee
and the following commutativity relation
\be
A \wedge B = (-1)^{pq} B \wedge A.
\ee
Thus, for example, 
\be
T^{I} = T^{I}{}_{\mu \nu} = T^{I}{}_{MN} e^{M}_{\mu} e^{N}_{\nu} = \frac{1}{2} T^{I}{}_{MN} e^{M} \wedge e^{N}.
\ee
Since the wedge product acts on spacetime indices only, it acts on the base manifold and not
on the internal fiber structure. 

With this in mind, the covariant derivative only acts on 
internal indices as follows:
\ba
D_{(\omega)} A^{KL} &:=& dA^{KL} + \omega^{KM} \wedge A_{M}{}^{L} + \omega^{LM} \wedge A^{K}{}_{M},
\nonumber \\
\\
D_{(\omega)} A_{KL} &:=& dA_{KL} - \omega_{K}{}^{M} \wedge A_{ML} - \omega_{L}{}^{M} \wedge A_{KM},
\nonumber \\
\ea
where the exterior derivative operator $d$ acts on spacetime indices only, namely
\be
dA^{KL} := 2 \partial_{[\mu} A^{KL}{}_{\nu]}.
\ee

From the anticommutator of covariant derivatives, one can define the curvature tensor associated with the spin connection
\be
F^{KL} = d\omega^{KL} + \omega^{K}{}_{M} \wedge \omega^{ML}.
\ee
With this definition at hand, one can easily show by direct computation that 
\be
\delta_{\omega} F^{IJ} = D_{(\omega)} \delta \omega^{IJ}.
\ee

We choose here to work with a spin connection that is internally compatible. In other words, we demand $D_{(\omega)} \eta^{IJ} = 0$, which then forces the spin connection to be fully antisymmetric on its internal indices $\omega^{(IJ)} = 0$. From this connection and the tetrad, one can also construct the torsion tensor defined as 
\be
\label{def-T}
T^{I} := D_{(\omega)} e^{I} = de^{I} + \omega^{I}{}_{M} \wedge e^{M},
\ee
which is equivalent to $T^{I}{}_{\mu \nu} = 2 D_{[\mu} e_{\nu]}^{I}$, or when spacetime indices are reinstated  
\be
\label{def-T2}
T^{\sigma}{}_{\mu \nu} = 2 \Gamma^{\sigma}_{[\mu \sigma]}.
\ee 
Note that internal metric compatibility is not equivalent to a torsion-free condition.

The contorsion tensor can be obtained from the definition of the torsion tensor.  We thus split the spin connection into a symmetric and tetrad compatible piece $\Gamma^{I}{}_{J}$ and an antisymmetric piece $C^{I}{}_{J}$, called the contorsion. The definition of the torsion tensor $D_{(\omega)} e^{I} = T^{I}$ then imposes 
\be
T^{I} = C^{I}{}_{J} \wedge e^{J},
\ee
or simply $T^{I}{}_{PQ} = -2 C^{I}{}_{[PQ]}$. These equations can be inverted to find
\be
C_{IJK} = - \frac{1}{2} \left(T_{IJK} + T_{JKI} + T_{KJI} \right).
\ee
Note that the contorsion is fully antisymmetric on its first two indices, while the torsion tensor is fully antisymmetric on its last two indices.  Also note that Eq.~\eqref{def-T2} can be obtained by converting Eq.~\eqref{def-T} to spacetime indices and using the transformation law from spin to spacetime connection established by $D_{(\Gamma)} e^{I} = 0$ (this relation is sometimes referred to as ``the tetrad postulate''). 

With the contorsion tensor, we can now express the curvature tensor in terms of the Riemann tensor $R^{IJ}$ and terms proportional to the contorsion
\be
F^{IJ} = R^{IJ} + D_{(\Gamma)} C^{IJ} + C^{I}{}_{M} \wedge C^{MJ},
\ee
where $D_{(\Gamma)}$ is the connection compatible with the symmetric connection. One can also check that the Bianchi identities in first order form become
\be
D_{(\omega)} T^{I} = R^{I}{}_{K} \wedge e^{K},
\qquad
D_{(\omega)} R^{IJ} = 0. 
\ee

Finally, it is sometimes useful to control the expression of the volume form in the first-order formalism. This quantity is given by
\be
\tilde{\sigma} := \sqrt{-g} \; d^{4}x =  \frac{1}{4!} \epsilon_{IJKL} e^{I} e^{J} e^{K} e^{L}
\ee
and it allows one to rewrite the contraction of the Levi-Civit\`{a} tensor with tetrad vectors in terms of $e$. 

\section{Other useful formulae}

In this appendix we present a compendium of other useful formulae, where the first expression corresponds to suppressed spacetime indices, 
followed by a second expression with spacetime indices reinstated, but transformed to internal ones with the tetrad.

We begin with the torsion tensor
\ba
T^{I} &=& \frac{1}{2} \frac{1}{\bar{\gamma}^{2} + 1} \left[ \epsilon^{I}{}_{JKL} \partial^{L} \bar{\gamma} + \bar{\gamma} \delta^{I}_{[J} \partial_{K]} \bar{\gamma} \right] e^{J} \wedge e^{K},
\nonumber \\
T_{I J K} &=& \frac{1}{\bar{\gamma}^{2} + 1} \left[ \epsilon_{IJK}{}^{L}  \partial_{L} \bar{\gamma} + \bar{\gamma} \delta_{I[J} \partial_{K]} \bar{\gamma} \right].
\ea
and the contorsion tensor
\ba
C_{IJ} &=& - \frac{1}{2} \frac{1}{\bar{\gamma}^{2} + 1} \left( \epsilon_{IJKQ} e^{K} \partial^{Q} \bar{\gamma} - 2 \bar{\gamma} e_{[I} \partial_{J]} \bar{\gamma} \right),
\nonumber \\
C_{IJK} &=& - \frac{1}{2} \frac{1}{\bar{\gamma}^{2} + 1} \left[ \epsilon_{IJKQ} \partial^{Q} \bar{\gamma} - 2 \bar{\gamma} \delta_{K[I} \partial_{J]} \bar{\gamma} \right].
\ea
%


\begin{thebibliography}{18}
\expandafter\ifx\csname natexlab\endcsname\relax\def\natexlab#1{#1}\fi
\expandafter\ifx\csname bibnamefont\endcsname\relax
  \def\bibnamefont#1{#1}\fi
\expandafter\ifx\csname bibfnamefont\endcsname\relax
  \def\bibfnamefont#1{#1}\fi
\expandafter\ifx\csname citenamefont\endcsname\relax
  \def\citenamefont#1{#1}\fi
\expandafter\ifx\csname url\endcsname\relax
  \def\url#1{\texttt{#1}}\fi
\expandafter\ifx\csname urlprefix\endcsname\relax\def\urlprefix{URL }\fi
\providecommand{\bibinfo}[2]{#2}
\providecommand{\eprint}[2][]{\url{#2}}

\bibitem[{\citenamefont{Ashtekar and Lewandowski}(2004)}]{Ashtekar:2004eh}
\bibinfo{author}{\bibfnamefont{A.}~\bibnamefont{Ashtekar}} \bibnamefont{and}
  \bibinfo{author}{\bibfnamefont{J.}~\bibnamefont{Lewandowski}},
  \bibinfo{journal}{Class. Quant. Grav.} \textbf{\bibinfo{volume}{21}},
  \bibinfo{pages}{R53} (\bibinfo{year}{2004}), \eprint{gr-qc/0404018}.

\bibitem[{\citenamefont{Thiemann}(2001)}]{Thiemann:2001yy}
\bibinfo{author}{\bibfnamefont{T.}~\bibnamefont{Thiemann}}
  (\bibinfo{year}{2001}), \eprint[http://arXiv.org/abs]{gr-qc/0110034}.

\bibitem[{\citenamefont{Rovelli}(2004)}]{Rovelli:2004tv}
\bibinfo{author}{\bibfnamefont{C.}~\bibnamefont{Rovelli}}
\bibinfo{title}{Quantum Gravity}
  (\bibinfo{year}{2004}), \bibinfo{note}{Cambridge, UK: Univ. Pr., 455 p}.

\bibitem[{\citenamefont{Romano}(1993)}]{Romano:1991up}
\bibinfo{author}{\bibfnamefont{J.~D.} \bibnamefont{Romano}},
  \bibinfo{journal}{Gen. Rel. Grav.} \textbf{\bibinfo{volume}{25}},
  \bibinfo{pages}{759} (\bibinfo{year}{1993}), \eprint{gr-qc/9303032}.

\bibitem[{\citenamefont{Ashtekar}(1991)}]{Ashtekar:1991hf}
\bibinfo{author}{\bibfnamefont{A.}~\bibnamefont{Ashtekar}}
  (\bibinfo{year}{1991})\bibinfo{title}{Lectures on Non-Pertubative Canonical Gravity}, \bibinfo{note}{Singapore, Singapore: World
  Scientific, 334 p. (Advanced series in astrophysics and cosmology, 6)}.

\bibitem[{\citenamefont{Barbero}(1995)}]{Barbero:1994ap}
\bibinfo{author}{\bibfnamefont{J.~F.} \bibnamefont{Barbero}},
  \bibinfo{journal}{Phys. Rev.} \textbf{\bibinfo{volume}{D51}},
  \bibinfo{pages}{5507} (\bibinfo{year}{1995}), \eprint{gr-qc/9410014}.

\bibitem[{\citenamefont{Holst}(1996)}]{Holst:1995pc}
\bibinfo{author}{\bibfnamefont{S.}~\bibnamefont{Holst}},
  \bibinfo{journal}{Phys. Rev.} \textbf{\bibinfo{volume}{D53}},
  \bibinfo{pages}{5966} (\bibinfo{year}{1996}), \eprint{gr-qc/9511026}.

\bibitem[{\citenamefont{Immirzi}(1997)}]{Immirzi:1996di}
\bibinfo{author}{\bibfnamefont{G.}~\bibnamefont{Immirzi}},
  \bibinfo{journal}{Class. Quant. Grav.} \textbf{\bibinfo{volume}{14}},
  \bibinfo{pages}{L177} (\bibinfo{year}{1997}), \eprint{gr-qc/9612030}.

\bibitem[{\citenamefont{Perez and Rovelli}(2006)}]{Perez:2005pm}
\bibinfo{author}{\bibfnamefont{A.}~\bibnamefont{Perez}} \bibnamefont{and}
  \bibinfo{author}{\bibfnamefont{C.}~\bibnamefont{Rovelli}},
  \bibinfo{journal}{Phys. Rev.} \textbf{\bibinfo{volume}{D73}},
  \bibinfo{pages}{044013} (\bibinfo{year}{2006}), \eprint{gr-qc/0505081}.

\bibitem[{\citenamefont{Freidel et~al.}(2005)\citenamefont{Freidel, Minic, and
  Takeuchi}}]{Freidel:2005sn}
\bibinfo{author}{\bibfnamefont{L.}~\bibnamefont{Freidel}},
  \bibinfo{author}{\bibfnamefont{D.}~\bibnamefont{Minic}}, \bibnamefont{and}
  \bibinfo{author}{\bibfnamefont{T.}~\bibnamefont{Takeuchi}},
  \bibinfo{journal}{Phys. Rev.} \textbf{\bibinfo{volume}{D72}},
  \bibinfo{pages}{104002} (\bibinfo{year}{2005}), \eprint{hep-th/0507253}.

\bibitem[{\citenamefont{Jordan}(1959)}]{Jordan:1959eg}
\bibinfo{author}{\bibfnamefont{P.}~\bibnamefont{Jordan}}, \bibinfo{journal}{Z.
  Phys.} \textbf{\bibinfo{volume}{157}}, \bibinfo{pages}{112}
  (\bibinfo{year}{1959}).

\bibitem[{\citenamefont{Brans and Dicke}(1961)}]{Brans:1961sx}
\bibinfo{author}{\bibfnamefont{C.}~\bibnamefont{Brans}} \bibnamefont{and}
  \bibinfo{author}{\bibfnamefont{R.~H.} \bibnamefont{Dicke}},
  \bibinfo{journal}{Phys. Rev.} \textbf{\bibinfo{volume}{124}},
  \bibinfo{pages}{925} (\bibinfo{year}{1961}).

\bibitem[{\citenamefont{Will}(2006)}]{lrr-2006-3}
\bibinfo{author}{\bibfnamefont{C.~M.} \bibnamefont{Will}},
  \bibinfo{journal}{Living Reviews in Relativity} \textbf{\bibinfo{volume}{9}}
  (\bibinfo{year}{2006}),
  \urlprefix\url{http://www.livingreviews.org/lrr-2006-3}.

\bibitem[{\citenamefont{Jackiw and Pi}(2003)}]{Jackiw:2003pm}
\bibinfo{author}{\bibfnamefont{R.}~\bibnamefont{Jackiw}} \bibnamefont{and}
  \bibinfo{author}{\bibfnamefont{S.~Y.} \bibnamefont{Pi}},
  \bibinfo{journal}{Phys. Rev.} \textbf{\bibinfo{volume}{D68}},
  \bibinfo{pages}{104012} (\bibinfo{year}{2003}), \eprint{gr-qc/0308071}.

\bibitem[{\citenamefont{Alexander and Yunes}(2008)}]{Alexander:2008wi}
\bibinfo{author}{\bibfnamefont{S.}~\bibnamefont{Alexander}} \bibnamefont{and}
  \bibinfo{author}{\bibfnamefont{N.}~\bibnamefont{Yunes}}
	\bibinfo{journal}{Phys. Rev.} \textbf{\bibinfo{volume}{D77}},
  \bibinfo{pages}{124040}  (\bibinfo{year}{2008}), 

\bibitem[{\citenamefont{Armendariz-Picon
  et~al.}(1999)\citenamefont{Armendariz-Picon, Damour, and
  Mukhanov}}]{ArmendarizPicon:1999rj}
\bibinfo{author}{\bibfnamefont{C.}~\bibnamefont{Armendariz-Picon}},
  \bibinfo{author}{\bibfnamefont{T.}~\bibnamefont{Damour}}, \bibnamefont{and}
  \bibinfo{author}{\bibfnamefont{V.~F.} \bibnamefont{Mukhanov}},
  \bibinfo{journal}{Phys. Lett.} \textbf{\bibinfo{volume}{B458}},
  \bibinfo{pages}{209} (\bibinfo{year}{1999}), \eprint{hep-th/9904075}.

\bibitem[{grt()}]{grtensor}
\emph{\bibinfo{title}{{GRTensorII}}}, \bibinfo{note}{this is a package which
  runs within Maple but distinct from packages distributed with Maple. It is
  distributed freely on the World-Wide-Web from the address: {\tt
  http://grtensor.org}}.

\end{thebibliography}

\end{document}